\documentclass[aps,prd,superscriptaddress,floatfix,nofootinbib,twocolumn]{revtex4-1}
\usepackage{epsfig}
\usepackage{bm}
\usepackage{amssymb}
\usepackage{amsmath}
\usepackage{color}
\usepackage{subfigure}
\usepackage{feynmp}

\begin{document}

\title{Linde problem in Yang-Mills theory compactified on $\mathbb{R}^2 \times \mathbb{T}^2$}
\date{\today}

\author{Eduardo S.~Fraga}
\affiliation{Instituto de F\'\i sica, Universidade Federal do Rio de Janeiro, 
Caixa Postal 68528, Rio de Janeiro, RJ 21941-972, Brazil}
\author{Daniel Kroff}
\affiliation{Instituto de F\'{\i}sica Te\'{o}rica, Universidade Estadual Paulista, Rua Dr. Bento Teobaldo Ferraz,
271 -- Bloco II, 01140-070 S\~ao Paulo, SP, Brazil}
\author{Jorge Noronha}
\affiliation{Instituto de F\'{\i}sica, Universidade de S\~{a}o Paulo, C.P. 66318,
05315-970 S\~{a}o Paulo, SP, Brazil}

\begin{abstract}
We investigate the perturbative expansion in $SU(3)$ Yang-Mills theory compactified on $\mathbb{R}^2\times \mathbb{T}^2$ where the compact space is a torus $\mathbb{T}^2= S^1_{\beta}\times S^1_{L}$, with $S^1_{\beta}$ being a thermal circle with period $\beta=1/T$ ($T$ is the temperature) while $S^1_L$ is a circle with finite length $L=1/M$, where $M$ is an energy scale. A Linde-type analysis indicates that perturbative calculations for the pressure in this theory break down already at order $\mathcal{O}(g^2)$ due to the presence of a non-perturbative scale $\sim g \sqrt{TM}$. We conjecture that a similar result should hold if the torus is replaced by any other compact surface of genus one. 
\end{abstract}

\maketitle


\section{Introduction}

Although finite-temperature field theory provides the natural framework to describe thermodynamic properties and phase transitions in plasmas involving gauge fields, its perturbative realization faces dreadful obstacles produced by severe infrared divergences in the gauge sector \cite{FTFT-books}. 

In the case of quantum chromodynamics (QCD) at finite temperature $T$, where the theory is defined in $\mathbb{R}^3 \times S^1_\beta$ and the compact direction in $S^1_\beta$ lies along the Euclidean time with period $\beta=1/T$, one can say that the domain of validity of the {\it naive}, plain perturbation expansion is tremendously restricted \cite{Freedman:1976ub,Shuryak:1977ut,Arnold:1994ps,Arnold:1994eb,Parwani:1994zz,Kajantie:2001hv} (see also Refs. \cite{Braaten:1995ju,Braaten:2002wi}). 

Over the years, this difficulty stimulated the development of techniques that reorganize the perturbative series, improving significantly the weak-coupling expansion (for reviews, see Refs. \cite{Blaizot:2001nr,Kraemmer:2003gd,Andersen:2004fp}). In particular, one can build an effective theory by using the separation of scales provided by $T$, $gT$, and $g^2T$, which is known as dimensionally reduced effective theory, or Electrostatic QCD (EQCD) \cite{Appelquist:1981vg,Kajantie:1995dw,Braaten:1995cm}. The pressure, for instance, is currently known up to ${\mathcal O}(g^6\ln g)$ at high temperatures and at most moderate chemical potentials $\mu_B\leq 10\,T$ \cite{Kajantie:2002wa,Vuorinen:2003fs,Hietanen:2008tv}. Alternatively, one can resort to the hard thermal loop (HTL) framework \cite{Braaten:1989mz,Braaten:1990az,Braaten:1991gm} (see Ref. \cite{Haque:2014rua} for recent 
results)\footnote{In a very recent development \cite{Kurkela:2016was}, these two approaches are combined in the treatment of cool quark matter: the zero Matsubara mode sector is treated via EQCD while the soft non-zero modes are resummed using HTL.}.

Nevertheless, at order $\mathcal{O}(g^{6})$ in the gauge coupling of a non-Abelian gauge theory at finite temperature, it is well-known that perturbation theory breaks down due to infrared divergences in the magnetic sector, the notorious Linde problem \cite{Linde:1980ts,Gross:1980br}. Therefore, to implement EQCD one is then obliged to match the coefficients of the order $\mathcal{O}(g^{6})$ to computations in lattice QCD in three dimensions \cite{Hietanen:2008tv,Hietanen:2004ew,Hietanen:2006rc,DiRenzo:2006nh}.

The compactification brought about by the finite temperature framework is the key ingredient responsible for the Linde problem since the static large distance behavior of the gauge theory at high $T$ is the
same as in 3-dimensional gauge theory, which confines at the scale $\sim g^2 T$. On the other hand, at zero temperature and finite density the Fermi sea does not produce dimensional reduction and this infrared problem is absent \cite{Pisarski:1998nh,Son:1998uk}. Therefore, it is interesting to investigate how Linde's analysis may be modified when thermal Yang-Mills theory is subjected to an additional compactification along a spatial direction.  

In this note we investigate the case of a pure glue $SU(3)$ plasma on a torus given by $\mathbb{R}^2 \times S^1_\beta\times S^1_L$, with $L=1/M$ being the length of the compactified spatial direction. By construction, this system is symmetric under the mapping $T \to M$ and $M \to T$, which we denominate {\it radius exchange symmetry}. Here we show that in this theory Linde's problem is much more severe and any perturbative calculation breaks down already at order $\mathcal{O}(g^2)$ due to the presence of a non-perturbative scale $\sim g \sqrt{TM}$. Therefore, observables in this theory (such as the pressure or screening masses) can only be computed non-perturbatively even when the gauge coupling is arbitrarily small. 

Such a setup may be useful to study some aspects of deconfinement and, in fact, a double trace deformation \cite{Pisarski:2006hz} of this theory in the limit of large number of colors was used in \cite{Simic:2010sv} to understand the interplay between the color electric and magnetic sectors in the deconfinement phase transition \cite{Liao:2006ry,Chernodub:2006gu}. Also, additional motivation to consider thermal non-Abelian gauge theories in spacetimes of different topology comes from the well-known study performed in \cite{Aharony:2003sx}, which considered large $N$ 4-dimensional thermal gauge theories in $S^3 \times S^1$. We shall see in the following that the severe infrared issues found here do not appear in that case.

This paper is organized as follows. In Section \ref{oldlinde} we summarize the main aspects behind Linde's original argument. In Section \ref{newlinde} we discuss the relevant properties of the perturbative Yang-Mills 
plasma on $\mathbb{R}^2 \times S^1_\beta\times S^1_L$ and the different regimes of the theory including the full breakdown of perturbation theory in this context. In Section \ref{conclusion} we present our final remarks 
and outlook.

\section{Linde's argument for thermal Yang-Mills theory}
\label{oldlinde}

In his seminal 1980 paper \cite{Linde:1980ts}, Linde argued that the perturbative study of Yang-Mills theory defined in $\mathbb{R}^3\times S^1_\beta$ is extremely problematic due to severe infrared divergences. Notably, perturbation theory for the pressure breaks down at order $\mathcal{O}(g^6)$. For completeness, we briefly review Linde's argument in the sequel.\\

\begin{figure}[ht]
\begin{center}
\resizebox*{!}{3.0cm}{\includegraphics{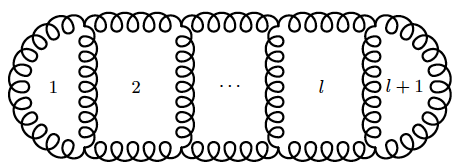}}
\end{center}
\vspace{-5mm}
\caption{\label{fig:linde_diagram} ($\ell+1$)-loop diagram for pure Yang-Mills theory.}
\end{figure}

Let us consider the contribution to the pressure from the $(\ell+1)$-loop diagram depicted in Fig. \ref{fig:linde_diagram}. The leading infrared (IR)
behavior comes from taking the Matsubara zero mode for every line; all other modes act as effective IR regulators. For the argument, it is sufficient to estimate its contribution to the pressure using a simple power counting strategy \cite{Linde:1980ts}. In this spirit, we can neglect its tensorial structure and write this zero mode contribution schematically in the form
\begin{equation}\label{eq:linde_estimate}
  g^{2\ell} \left( T \int d^3 k\right)^{\ell+1} \frac{k^{2\ell}}{(k^2)^{3\ell}}\,,
\end{equation}
where $K^\mu = \left(\omega_{n},\vec{k}\right)$ is the gluon 4-momentum and $\omega_{n}=2\pi T n$ is the Matsubara frequency. The origin of each factor is easy to trace: vertices contribute with $g \times k$, lines with a propagator $(k^2)^{-1}$, and loops with an integral.

In order to estimate the contribution from this diagram, we take $T$ as an ultraviolet cutoff since it would naturally arise if we were to sum over all Matsubara frequencies due to the presence of statistical factors. Also, as the diagrams are potentially IR  divergent, one cannot take arbitrarily soft modes into account; we only integrate over momentum above a lower threshold $a$.

The diagrams are IR regular for $\ell < 3$ and divergent for $\ell \geq 3$. Namely, the dominant behavior for different values of $\ell$ are
\cite{Linde:1980ts,FTFT-books}
\begin{subequations}\label{eq:linde_result}
\begin{alignat}{3}
&\sim g^{2\ell} T^4                \quad &&\textrm{for } \ell < 3, \label{subeq:linde_result_reg}	\\
&\sim g^6 T^4 \log \frac{T}{a}        \quad &&\textrm{for } \ell = 3, \label{subeq:linde_result_log}	\\
&\sim g^6 T^4 \left(\frac{g^2 T}{a}\right)^{\ell-3}   \quad &&\textrm{for } \ell > 3. \label{subeq:linde_result_pl}
\end{alignat}
\end{subequations}

In perturbation theory, the coupling $g$ and the scale $T$ provide a natural hierarchy of energy scales:
$T > gT > g^2 T > \cdots$. Using such hierarchy as a guideline, one can push $a$ deeper and deeper towards the IR region. Eq.\ \eqref{subeq:linde_result_pl} shows that when $a$ reaches $g^2 T$ all diagrams with $\ell > 3$ contribute at $\mathcal{O}(g^6)$. In other words, 
perturbation theory breaks down since infinitely many diagrams have to be considered at a finite order, even if $g$ is taken to be arbitrarily small.


The reasoning above ignores the possibility of screening masses being dynamically generated and, as a matter of fact, they are present
in thermal Yang-Mills theory \cite{FTFT-books}. In this context, a screening mass would work as a natural IR regulator, essentially playing the role of $a$ in Eq.\ \eqref{eq:linde_result}. In the color electric sector, IR modes are screened as $a = m_{\textrm{el}} \sim gT$. On the other hand, from Eq.\ \eqref{subeq:linde_result_pl} one can see that a color magnetic mass $a= m_{\textrm{mag}} \sim g^2T$ makes all loops $\ell > 3 $ contribute to $\mathcal{O}(g^6)$, which is interpreted as the breakdown of the perturbative expansion. This is the so-called Linde problem of thermal Yang-Mills theory.


\section{Linde problem on the torus}
\label{newlinde}

Now we consider pure glue $SU(3)$ Yang-Mills theory in $\mathbb{R}^2 \times S^1_\beta\times S^1_L$. We define our coordinates as $x^\mu = (x,y,\tau,\xi)$ where $(x,y)$ corresponds to $\mathbb{R}^2$ and $\tau \in [0,\beta=1/T]$ and $\xi = [0,L=1/M]$ parametrize the torus. We note that the partition function is periodic in $\tau$ {\it and} $\xi$ and all observables in this theory should be invariant under radius exchange symmetry, i.e., $M \Longleftrightarrow T$. The Fourier decomposition of the Yang-Mills field is given by
\begin{equation}
A_\mu(x,y,\tau,\xi) = \sum_{m,n=-\infty}^{\infty}A_\mu^{(m,n)} (x,y) e^{i n \tau/\beta} e^{i m \xi /L}
\end{equation}
and, due to the presence of two compact dimensions, this system can be formally seen as Kaluza-Klein like tower of two-dimensional Yang-Mills theories coupled to two adjoint scalars for each one of the winding modes on the torus. We denote $A_\tau^{(m,n)}(x,y) = \phi^{(m,n)}(x,y)$ and $A_\xi^{(m,n)}(x,y)= \Phi^{(m,n)}(x,y)$, where $\phi$ and $\Phi$ represent these two adjoint scalars. As discussed in \cite{Simic:2010sv}, this system has global $(\mathbb{Z}_3)_\beta \times (\mathbb{Z}_3)_L$ center symmetry and two order parameters given by the Wilson lines on the torus.

One may consider the behavior of this theory in certain limits of the energy scales $T$ and $M$:
\begin{itemize}
\item $T,M \to \infty$ (dimensional reduction): the adjoint scalars acquire a large mass and decouple from the low-energy effective theory, which becomes the exactly solvable two-dimensional Yang-Mills theory for the massless gluons along the two non-compact directions \cite{Gross:1980he};

\item $T \to 0$, $M$ finite (or $M \to 0$, $T$ finite): One of the compact dimensions unwinds and the low-energy effective theory becomes three-dimensional Yang-Mills theory coupled to an adjoint scalar of mass $\sim g T$ (or $\sim gM$).
\end{itemize}

\begin{figure}[ht]
\begin{center}
\resizebox*{!}{5.0cm}{\includegraphics{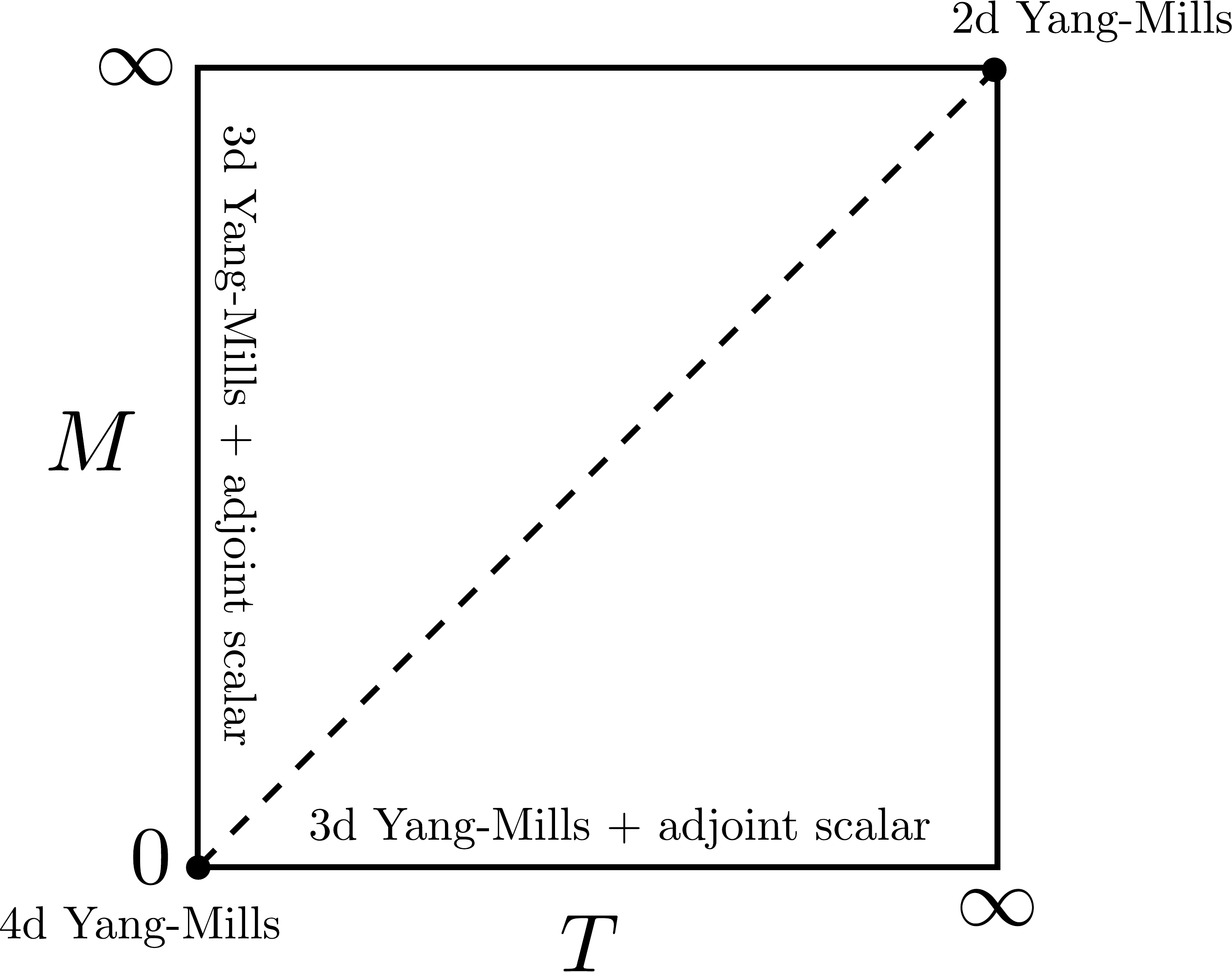}}
\end{center}
\vspace{-5mm}
\caption{Behavior of the theory in the $M-T$ plane.}
\label{diagrama}
\end{figure}

These scenarios are illustrated in Fig.\ \ref{diagrama}. When $M \to 0$, $T \neq 0$ the two-point function of the color field tensor ${\rm Tr}F_{\mu\nu}F^{\mu\nu}$ with components along the non-compact directions exhibits the usual screening for large spatial separations while when $M, T \to \infty$ this correlator becomes non-propagating as in 2-dimensional Yang-Mills theory \cite{Gross:1980he}. 

Now we can address Linde's problem in $\mathbb{R}^2 \times S^1_\beta\times S^1_L$. As before, we estimate in a power counting scheme the dominant IR contribution of the Linde diagrams, shown in Fig.\ \ref{fig:linde_diagram}. In the present case there is not one but two compactified dimensions and, thus, in order to get the leading IR contribution we must take the zero modes associated with each circle for every line in the loop diagram. The equivalent of Eq.\ \eqref{eq:linde_estimate} is then
\begin{equation}
  g^{2\ell} \left( M T \int d^2 k\right)^{\ell+1} \frac{k^{2\ell}}{(k^2)^{3\ell}}.
\end{equation}

As in Section \ref{oldlinde}, if we were to sum over all modes a natural UV cutoff would arise. Such a hard scale is given as a function of the two energy scales related to the two compact directions, which we denote by $ f(T, M)$. Its exact dependence on $M$ 
and $T$ is not important for our argument though it must satisfy the radius exchange symmetry, i.e., $f(M, T) = f(T, M)$.

Once again, due to the potential IR divergences of the Linde diagrams, we only integrate over modes above a certain IR scale $a$. The dominant IR behavior for different values of $\ell$ can be readily estimated:
\begin{subequations}\label{eq:linde_result_torus}
\begin{alignat}{2}
&\sim g^2 M^2T^2 \log \frac{f(M,T)}{a}        \quad &&\textrm{for } \ell = 1, \label{subeq:linde_result_torus_log}	\\
&\sim g^2 M^2T^2 \left(\frac{g \sqrt{M T}}{a}\right)^{2\ell-2}   \quad &&\textrm{for } \ell > 1. \label{subeq:linde_result_torus_pl}
\end{alignat}
\end{subequations}
 As the problem now has two typical scales, one can no longer build a unique hierarchy of energy scales with the aid of the coupling constant. In fact, there are infinitely many hierarchies at our disposal, one for each possible combination of $M$ and $T$ with the proper dimension. Nevertheless, for scales where $a \sim g \sqrt{M T}$, all Linde diagrams with 3 or more loops contribute at $\mathcal{O}(g^2)$, which indicates the breakdown of perturbation theory. Thus, the presence of a second compactified dimension renders the Linde problem in gauge theories even more severe. The only analytically computable case is the Stefan-Boltzmann (ideal gas) limit, which gives the following result for the pressure using standard finite temperature techniques \cite{FTFT-books} (and $N=3$) 
\begin{eqnarray}
&&P_{SB} = \frac{2\pi^2}{15}  \left(T^4 + M^4    \right) + \frac{2\pi^2}{9} T^2 M^2 + 16\,T^2 M^2 \nonumber \\ &\times& \sum_{n=1}^\infty \frac{1}{n^2}\left[\frac{e^{2n\pi \beta M}}{\left( e^{2n\pi \beta M}-1  \right)^2}+\frac{e^{2n\pi/ (\beta M)}}{\left( e^{2n\pi/ (\beta M)}-1  \right)^2}    \right],\nonumber \\
\label{psb}
\end{eqnarray}
where radius exchange symmetry is manifest (also, note that \eqref{psb} reduces to the well-known result $8 \pi^2 T^4/45$ when $M\to 0$).

The fact that IR fluctuations have become stronger should be expected given the particular type of compact space we considered and the presence of an additional zero mode\footnote{We thank R.~Pisarski for pointing this out to us.} in the torus in comparison to the usual case of YM in $\mathbb{R}^3 \times S^1_\beta$. Also, we note that the zero modes in the compact directions, though constant, have non-trivial commutators. A thorough discussion about these modes is, however, beyond the scope of the present note. 
 
In this regard, we would like to point out that there is an important difference between the case considered by Aharony {\it et al.} in Ref.\ \cite{Aharony:2003sx} and the one we
address in this paper: the eigenmodes of the Laplacian operator. In our case, $\mathbb{R}^2\times S^1_\beta \times S^1_L$, the eigenmodes are, along all four directions, Fourier modes (plane waves) and the corresponding eigenvalues are simply the square of arbitrary real numbers and the square of the Matsubara frequencies for 
the non-compact and compactified directions, respectively. Therefore, the propagator is bound to diverge
when the modes with vanishing eigenvalues are considered, which introduces IR divergences in the computation of Feynman diagrams. On the other hand, when embedding the theory in $S^3 \times S^1$ as in \cite{Aharony:2003sx}, the eigenmodes are one Fourier mode along the time direction and the 3-dimensional generalization of the vector spherical harmonics over the 3-sphere. The eigenvalues related to the Fourier mode are, once more, the square of the Matsubara frequencies but the one related to the spherical harmonics, which the authors of Ref.\ \cite{Aharony:2003sx} call $\Delta^2$, can be shown to be a positive integer. In other words, the eigenvalues related to the eigenmodes living on $S^3$ never vanish and, thus, work as a natural IR regulator -- the propagator never diverges even when the zero-Matsubara mode is taken into account. Thus, one can see that there is a link between the topology of the spacetime within which thermal gauge theories are embedded and the fate of perturbation theory.

In the dimensional reduction limit, i.e., $T,M \to \infty$, the system becomes effectively Yang-Mills theory in two dimensions \cite{Gross:1980he}. In this case, one can write the partition function purely in terms of the field strengths $F_{{\mu\nu}}$ (Bianchi's constraint is trivial in two dimensions) \cite{Halpern:1978ik}, with a simple quadratic action $\sim 1/g^{2}$ whose field strengths fluctuate on the plane independently from one another. The pressure of this system can be solved exactly \cite{Gross:1980he} and it contains a term that goes as $\sim 1/g^{2}$  and, thus, it cannot be simply expanded in perturbative powers of $g$ at weak coupling. We believe this may be the root behind the failure of naive perturbation theory already at order $g^{2}$ found here for Yang-Mills theory on  $\mathbb{R}^2 \times S^1_\beta\times S^1_L$. 

Additionally, we remark that 2-dimensional Yang-Mills theory would also appear as a limit of the original 4-dimensional theory if the torus would be replaced by any compact surface with genus one, though the explicit construction of the Linde problem in this case is beyond the scope of this paper.


\section{Conclusion and outlook}
\label{conclusion}

Perturbative expansions at finite temperature are plagued with IR divergences whenever massless bosonic fields are present \cite{FTFT-books}.
This situation is even more problematic in the case of thermal Yang-Mills theory since the Linde problem essentially makes naive perturbation theory meaningless 
beyond $\mathcal{O}(g^6)$.

Thermal field theories, in the imaginary-time formalism, are set in an Euclidean space-time with a compact time direction and, in this context, all thermal effects are ultimately encoded in the structure of the underlying space-time (e.g., $\mathbb{R}^3 \times S^1_\beta$). In order to better understand the role played by compactification in IR problems of non-Abelian gauge theories, we included a second compact dimension and 
analyzed the behavior of Yang-Mills theory, providing an extension of Linde's argument for the case where the compact part of space-time is the torus $\mathbb{T}^2=S^1_\beta\times S^1_L$.

Our study shows that the Linde problem in this case becomes much more severe, as it already emerges at $\mathcal{O}(g^2)$ for the pressure. This indicates that the perturbative expansion in Yang-Mills theory on $\mathbb{R}^2 \times S^1_\beta\times S^1_L$ faces important
limitations since the lowest order correction to any physical observable will necessarily have a non-perturbative contribution even at arbitrarily small coupling. 
However, this system could be readily studied on the lattice and it would be interesting 
to investigate the phase diagram of this theory, depicted in Fig.\ \ref{diagrama}. One could compute on the lattice the glueball correlator and see its behavior changing from the well-known description in terms of screening masses when $L \to \infty$ to the finite $L$ scenario addressed in this paper. 

Finally, it would be interesting to see if the breakdown of perturbation theory induced by IR divergences in Yang-Mills theory indeed has a topological character. One could check if other compact surfaces with genus one (which are then topologically equivalent to the torus considered here) produce the same qualitative results for the pressure. While we cannot rigorously prove it at this time, we conjecture that this is going to be the case because the dimensional reduction argument discussed here, which leads to 2-dimensional Yang-Mills theory, should also hold. Additionally, compactifications of Yang-Mills theory on $\mathbb{R}^2\times \mathbb{M}^2$, where $\mathbb{M}^2$ is a compact surface with genus $>1$, may yet reveal other features that are not present in the simple torus example considered here. 

\section*{Acknowledgements}

The authors are indebted to R.~D.~Pisarski and A.~Vuorinen for useful comments. J.~N. thanks A.~Dumitru for discussions. 
This work was supported by  CAPES, CNPq, FAPERJ, and FAPESP.



\end{document}